\documentstyle[prd,aps,psfig,floats,preprint]{revtex}
\begin{document}

\def\beq{\begin{equation}}
\def\eeq{\end{equation}}
\def\ber{\begin{eqnarray}}
\def\eer{\end{eqnarray}}
\def \lleq {\lower0.9ex\hbox{ $\buildrel < \over \sim$} ~}
\def \ggeq {\lower0.9ex\hbox{ $\buildrel > \over \sim$} ~}
\def\l{\Lambda}

\def\lsim{\
  \lower-1.5pt\vbox{\hbox{\rlap{$<$}\lower5.3pt\vbox{\hbox{$\sim$}}}}\ }
\def\gsim{\
  \lower-1.5pt\vbox{\hbox{\rlap{$>$}\lower5.3pt\vbox{\hbox{$\sim$}}}}\ }

\def\apj{{Astroph.\@ J.\ }}
\def\mn{{Mon.\@ Not.@ Roy.\@ Ast.\@ Soc.\ }}
\def\asta{{Astron.\@ Astrophys.\ }}
\def\aj{{Astron.\@ J.\ }}
\def\prl{{Phys.\@ Rev.\@ Lett.\ }}
\def\prd{{Phys.\@ Rev.\@ D\ }}
\def\nucp{{Nucl.\@ Phys.\ }}
\def\nat{{Nature\ }}
\def\plb {{Phys.\@ Lett.\@ B\ }}
\def \jetpl {JETP Lett.\ }
\def\etal{{\it et al.}}

\title{Braneworld models of dark energy}
\author{Varun Sahni$^a$ and Yuri Shtanov$^b$}
\address{$^a$Inter-University Centre for Astronomy and Astrophysics, Post Bag 4,
Ganeshkhind, Pune 411~007, India \\
$^b$Bogolyubov Institute for Theoretical Physics, Kiev 03143, Ukraine}
\maketitle

\begin{abstract}
We explore a new class of braneworld models in which the scalar curvature of
the (induced) brane metric contributes to the brane action. The scalar
curvature term arises generically on account of one-loop effects induced by
matter fields residing on the brane. Spatially flat braneworld models can enter
into a regime of accelerated expansion at late times. This is true even if the
brane tension and the bulk cosmological constant are tuned to satisfy the
Randall--Sundrum constraint on the brane. Braneworld models admit a wider range
of possibilities for dark energy than standard LCDM. In these models the
luminosity distance can be both smaller and larger than the luminosity distance
in LCDM. Whereas models with $d_L \leq d_L(\rm LCDM)$ imply $w = p/\rho \geq
-1$ and have frequently been discussed in the literature, models with $d_L >
d_L(\rm LCDM)$ have traditionally been ignored, perhaps because within the
general-relativistic framework, the luminosity distance has this property {\em
only if\/} the equation of state of matter is strongly negative ($w < -1$).
Within the conventional framework, `phantom energy' with $w < -1$ is beset with
a host of undesirable properties, which makes this model of dark energy
unattractive. Braneworld models, on the other hand, have the capacity to endow
dark energy with exciting new possibilities (including $w < -1$) without
suffering from the problems faced by phantom energy. For a subclass of
parameter values, braneworld dark energy and the acceleration of the universe
are {\em transient\/} phenomena. In these models, the universe, after the
current period of acceleration, re-enters the matter-dominated regime so that
the deceleration parameter  $q(t) \to 0.5$ when $t \gg t_0$, where $t_0$ is the
present epoch. Such models could help reconcile an accelerating universe with
the requirements of string/M-theory.
\end{abstract}

\section{Introduction}

One of the most sensational discoveries of the past decade is that our universe
is currently accelerating. This observation finds support in the luminosity
measurements of high-redshift supernovae \cite{sn}, measurements of
degree-scale anisotropies in the cosmic microwave background \cite{wmap} and,
indirectly, in the observations of gravitational clustering \cite{2df}.
Although a cosmological constant appears to satisfy all current observations,
the formidable fine-tuning difficulties which accompany a non-evolving
$\l$-term have prompted theorists to investigate alternative, evolving forms of
dark energy. Theoretical models of dark energy usually involve matter fields
possessing unusual properties such as a negative equation of state. Popular
dark-energy models include scalar-field-based `quintessence' models, models
based on quantum particle production, Chaplygin gas etc.\@ (see
\cite{ss00,sahni01,rp02} for recent reviews). In this paper, we examine a
radically different mechanism for the late time acceleration of the universe in
which dark energy effectively emerges from the gravity sector and not the
matter sector of the theory.

%Although the study of higher dimensional `braneworld models' is the subject of
%intense scientific activity, several issues relating to the cosmological
%implications of these models have not yet been sufficiently well investigated.
%In this paper, we would like to address the issue of the cosmological dark
%energy within the context of the braneworld scenario. We consider a general
%model which is purely gravitational in the bulk but includes the scalar
%curvature term in the action for the brane. The presence of the curvature term
%is essential for the braneworld dynamics. We restrict ourselves to the case
%where the brane is a boundary of the five-dimensional bulk, which is equivalent
%to the situation with $Z_2$ isometry of reflection with respect to the brane
%embedded in the bulk.  Thus, the model under consideration is described by four
%parameters, namely, the bulk and brane gravitational and cosmological
%constants.

We study a higher-dimensional cosmology in which the observable universe is a
four-dimensional `brane' embedded in a five-dimensional bulk. Models of this
kind appeared after the seminal paper by Randall \& Sundrum (RS) \cite{RS}.
Subsequent intensive investigation showed that cosmology based on the RS model
exhibits departure from standard general-relativistic behaviour only at very
early cosmological times \cite{BDL,SMS,brane1}.  The model which we study in
this paper generalizes the RS model in that, besides the brane and bulk
cosmological constants, it also includes the scalar curvature term in the
action for the brane.
%In comparison, the braneworld model of Dvali, Gabadadze,
%and Porrati (DGP) \cite{DGP} contains the curvature term in the action for the
%brane but has bulk and brane cosmological constants expressly set equal to
%zero.
In an interesting recent development, Deffayet, Dvali, and Gabadadze (DDG)
\cite{DDG}
demonstrated that the presence of the scalar curvature term in the action for
the brane can lead to a late-time acceleration of the universe even in the
absence of any material form of dark energy \cite{DDG,DGP,Deffayet}.
%or cosmological constant.
The main difference between our model and the DDG model is that, in addition to
the scalar curvature term in the action for the brane, we also include the
brane and bulk cosmological constants.
%It is important to mention though, that in contrast to our model,
(The bulk and brane cosmological constants were set equal to zero
in the DDG model.)
%Introduction of the curvature term in the action for the brane allows one to
%have braneworld models whose behaviour depart from the standard one at {\em
%late\/} cosmological times \cite{CH,Shtanov1}, and to account for the observed
%acceleration of the universe even with zero bulk and brane cosmological
%constants \cite{Deffayet,DDG}.
%
%In this paper, we consider a general braneworld model
%which includes both the curvature
%term in the action for the brane as well as
%the cosmological constants in the bulk and brane respectively.

The presence of bulk and brane cosmological constants in our braneworld (in
addition to the brane curvature term) leads to several qualitatively new
features that distinguish our model from the DDG model as well as from the
scalar-field-based `quintessence' models. An important property of our
braneworld universe is that the (effective) equation of state of dark energy
can be $w < - 1$. (The DDG braneworld model and quintessence models have $w >
-1$.) In addition, the acceleration of the universe in our braneworld scenario
can be a {\em transient\/} phenomenon. We find that a class of braneworld
models accelerate during the present epoch but revert back to matter-dominated
expansion at late times. A {\em transiently accelerating\/} braneworld does not
possess an event horizon and could reconcile a currently accelerating universe
with the demands of string/M-theory.

%As a result braneworld cosmology has the
%following qualitatively new features: (i)~The
%luminosity distance $d_L(z)$ in a particular class of braneworld models can
%{\em exceed\/} that in the LCDM model leading to an effective
%equation of state of dark energy $w < - 1$. (ii)~Braneworld models can enter
%into a regime of accelerated expansion at late times even if the brane and bulk
%cosmological constants are tuned to satisfy the Randall--Sundrum constraint on
%the brane.  In this case, braneworld dark energy and the acceleration of the
%universe are {\em transient\/} phenomena, and the universe, after the current
%period of acceleration, re-enters the matter-dominated regime.
%These two features are very interesting from an
% observational point of view
%and warrant serious attention.
%(We are not aware of any other braneworld model that can give these results.)

Our paper is organised as follows. After describing the braneworld model in the
next section, in Sec.~\ref{vacua} we proceed to the case of vacuum braneworlds,
that is, solutions with zero stress-energy tensor of matter.  We derive a
general expression for the effective cosmological constant in such models and
describe some classes of its symmetric solutions, including {\em empty\/}
static homogeneous and isotropic universes.  We also present simple expressions
for a static universe filled with matter.  In Sec.~\ref{cosmology} we study the
cosmological evolution of the braneworld. We shall show that braneworld models
have properties similar to the observed (accelerating) universe for a large
region of parameter space. For instance, the model under consideration can
mimic the evolution of a FRW universe in which dark energy has
pressure--to--energy-density ratio $p / \rho$ both larger {\em as well as
smaller\/} than the critical value of $-1$. For a subclass of parameter values,
braneworld dark energy and the acceleration of the universe are {\em
transient\/} phenomena so that the universe, after the current period of
acceleration, re-enters the matter-dominated regime.  In Sec.~\ref{final} we
formulate our conclusions.

\section{Basic equations} \label{basic}

In this paper, we consider the case where a braneworld is the timelike boundary
of a five-dimensional purely gravitational Lorentzian space (bulk), which is
equivalent to the case of a brane embedded in the bulk with $Z_2$ symmetry of
reflection with respect to the brane.  The theory is described by the action
\cite{CH,Shtanov1}
\begin{equation} \label{action}
S = \epsilon M^3 \left[\int_{\rm bulk} \left( {\cal R} - 2 \Lambda_{\rm b}
\right) - 2 \int_{\rm brane} K \right] + \int_{\rm brane} \left( m^2 R - 2
\sigma \right) + \int_{\rm brane} L \left( h_{ab}, \phi \right) \, .
\end{equation}
Here, ${\cal R}$ is the scalar curvature of the metric $g_{ab}$ in the
five-dimensional bulk, and $R$ is the scalar curvature of the induced metric
$h_{ab} = g_{ab} - n_a n_b$ on the brane, where $n^a$ is the vector field of
the inner unit normal to the brane, and the notation and conventions of
\cite{Wald} are used. The quantity $K = K_{ab} h^{ab}$ is the trace of the
symmetric tensor of extrinsic curvature $K_{ab} = h^c{}_a \nabla_c n_b$ of the
brane. The symbol $L (h_{ab}, \phi)$ denotes the Lagrangian density of the
four-dimensional matter fields $\phi$ whose dynamics is restricted to the brane
so that they interact only with the induced metric $h_{ab}$. All integrations
over the bulk and brane are taken with the natural volume elements $\sqrt{-
g}\, d^5 x$ and $\sqrt{- h}\, d^4 x$, respectively, where $g$ and $h$ are the
determinants of the matrices of components of the corresponding metrics in a
coordinate basis. The symbols $M$ and $m$ denote, respectively, the
five-dimensional and four-dimensional Planck masses, $\Lambda_{\rm b}$ is the
bulk cosmological constant, and $\sigma$ is the brane cosmological constant,
also called the brane tension.

The parameter $\epsilon = \pm 1$ in the first term of action (\ref{action})
reflects the possibility of different relative signs between the bulk and brane
terms in the action.  In the context of homogeneous Friedmann cosmology that
will be subject of this paper, the differential equations on the brane will not
depend on the sign of $\epsilon$.  However, different signs of $\epsilon$
result in different ways of embedding of a brane with one and the same induced
metric in the bulk space, which may be important for studying perturbations of
the theory.

The term containing the scalar curvature of the induced metric on the brane
with the coupling $m^2$ in action (\ref{action}) is often neglected in the
literature. However, this term is qualitatively essential for describing the
braneworld dynamics since it is inevitably generated as a quantum correction to
the matter action in (\ref{action}) --- in the spirit of an idea that goes back
to Sakharov \cite{Sakharov} (see also \cite{BD}). Note that the effective
action for the brane typically involves an infinite number of terms of higher
order in curvature (this was pointed out in \cite{CH} for the case of
braneworld theory; a similar situation in the context of the AdS/CFT
correspondence is described in \cite{HHR}).  In this paper, we retain only the
terms linear in curvature. The linear effects of the curvature term on the
brane were studied in \cite{DGP,CH2}, where it was shown that it leads to
four-dimensional law of gravity on sufficiently small scales.  For some recent
reviews of the results connected with the induced-gravity term in the brane
action, one may look into \cite{review}.

Action (\ref{action}) leads to the bulk being described by the usual Einstein
equation with cosmological constant:
\begin{equation} \label{bulk}
{\cal G}_{ab} + \Lambda_{\rm b} g_{ab} = 0 \, ,
\end{equation}
while the field equation on the brane is
\begin{equation} \label{brane}
m^2 G_{ab} + \sigma h_{ab} = \tau_{ab} + \underline{\epsilon M^3 \left(K_{ab} -
h_{ab} K \right)} \, .
\end{equation}
Here, $\tau_{ab}$ is the stress-energy tensor on the brane, and it stems from
the last term in action (\ref{action}).  In Eq.~(\ref{brane}) we have
underlined the term whose presence makes braneworld theory different from
general relativity. (It should perhaps be mentioned that the variation of
action (\ref{action}) which includes the Gibbons--Hawking surface term
$\displaystyle\int_{\rm brane}\!K$ leads to the Israel junction conditions
\cite{israel} on the brane, as demonstrated in \cite{CH,Shtanov1}.) From
(\ref{brane}) one can show that there exists a scale whose value determines the
domain in which general relativity is approximately valid \cite{DGP,CH2}.
Indeed, we can write
\begin{equation}
G_{ab} \sim r_1^{-2} \, , \quad K_{ab} \sim r_2^{-1} \, ,
\end{equation}
where $r_1$ and $r_2$ are the characteristic curvature radii of solution of
Eq.~(\ref{brane}). Then the ratio of the last term on the right-hand side of
Eq.~(\ref{brane}) to the first term on its left-hand side is of the order
$r_1^2 / r_2 \ell$, where $\ell = 2 m^2 /M^3$ is a convenient definition to be
used later. Thus, for $r_1^2 \ll r_2 \ell$, the last term on the left-hand side
of Eq.~(\ref{brane}) can be neglected, and general relativity is recovered.  In
typical situations, e.g., in cosmology, we often have $r_1 \sim r_2 \sim r$,
and general relativity becomes valid for $r \ll \ell$.  This fact was used in
\cite{DGP}, where the cosmological model with $\sigma = 0$ and $\Lambda_{\rm b}
= 0$ was considered. Therefore, an important distinguishing feature of the
braneworld models which we consider, is that they can show departure from
standard FRW behaviour at {\em late times\/} and on large scales. This property
should be contrasted with the braneworld cosmology based on the
Randall--Sundrum model \cite{RS} (with $m = 0$), in which nonstandard behaviour
is encountered at very early times \cite{BDL,SMS,brane1}.

By contracting the Gauss identity
\begin{equation}
R_{abc}{}^d = h_a{}^f h_b{}^g h_c{}^k h^d{}_j {\cal R}_{fgk}{}^j + K_{ac}
K_b{}^d - K_{bc} K_a{}^d
\end{equation}
on the brane and using Eq.~(\ref{bulk}), one obtains the `constraint' equation
\begin{equation} \label{constraint}
R - 2 \Lambda_{\rm b} + K_{ab} K^{ab} - K^2 = 0\, ,
\end{equation}
which, together with (\ref{brane}), implies the following closed scalar
equation on the brane:
\begin{equation} \label{closed}
M^6 \left( R - 2 \Lambda_{\rm b} \right) + \left( m^2 G_{ab} + \sigma h_{ab} -
\tau_{ab} \right) \left( m^2 G^{ab} + \sigma h^{ab} - \tau^{ab} \right) - {1
\over 3} \left( m^2 R - 4 \sigma + \tau \right)^2 = 0\, ,
\end{equation}
where $\tau = h^{ab} \tau_{ab}$.  One method for obtaining solutions of the
theory consists in first solving the scalar equation (\ref{closed}) on the
brane together with the stress-energy conservation equation, and then
integrating the Einstein equations in the bulk with the given data on the brane
\cite{SMS,Shtanov2}.

We note that Eq.~(\ref{closed}) describes the evolution on the brane in terms
of its intrinsic quantities, and that all homogeneous and isotropic
cosmological solutions for the brane can be obtained from this equation
\cite{Shtanov2}, as will also be done below.  All such solutions are embeddable
in the Schwarzschild-AdS five-dimensional bulk.  However, one cannot confine
oneself to this single equation in the general case, e.g., in studying
perturbations of the cosmological solution, where one must also solve equations
in the bulk imposing certain boundary and/or regularity conditions. From this
general viewpoint, Eq.~(\ref{closed}) cannot be regarded as complete: only
those of
its solutions are admissible which can be developed to regular solutions
in the bulk.  This illustrates the importance of the precise specification of
boundary and/or regularity conditions in the bulk for a complete formulation of
the brane-world theory \cite{Shtanov2}.

The gravitational equations in the bulk can be integrated by using, for
example, Gaussian normal coordinates, as described, e.g., in \cite{Shtanov3}.
Specifically, in the Gaussian normal coordinates $(x, y)$, where $x =
\{x^\alpha\}$ are the coordinates on the brane and $y$ is the fifth coordinate
in the bulk, the metric in the bulk is written as
\begin{equation}
d s^2_5 = dy^2 + h_{\alpha\beta} (x, y) dx^\alpha dx^\beta \, .
\end{equation}
Introducing also the tensor of extrinsic curvature $K_{ab}$ of every
hypersurface $y = {\rm const}$ in the bulk, one can obtain the following system
of differential equations for the components $h_{\alpha\beta}$ and
$K^\alpha{}_\beta$:
\begin{equation} \label{s}\begin{array}{lcl}
\displaystyle {\partial K^\alpha{}_\beta \over \partial y} &=& \displaystyle
R^\alpha{}_\beta - K K^\alpha{}_\beta - \frac16 \delta^\alpha{}_\beta \left( R
+ 2 \Lambda_{\rm b} + K^\mu{}_\nu K^\nu{}_\mu - K^2 \right) \medskip \\ &=&
\displaystyle R^\alpha{}_\beta - K K^\alpha{}_\beta - \frac23
\delta^\alpha{}_\beta \Lambda_{\rm b} \, ,
\end{array}
\end{equation}
\begin{equation} \label{metric}
{\partial h_{\alpha\beta} \over \partial y} = 2 h_{\alpha\gamma}
K^\gamma{}_\beta \, ,
\end{equation}
where $R^\alpha{}_\beta$ are the components of the Ricci tensor of the metric
$h_{\alpha\beta}$ induced on the hypersurface $y = {\rm const}$, $R =
R^\alpha{}_\alpha$ is its scalar curvature, and $K = K^\alpha{}_\alpha$ is the
trace of the tensor of extrinsic curvature. The second equality in (\ref{s}) is
true by virtue of the `constraint' equation (\ref{constraint}).  Equations
(\ref{s}) and (\ref{metric}) together with the `constraint' equation
(\ref{constraint}) represent the $4\!+\!1$ splitting of the Einstein equations
in the Gaussian normal coordinates. The initial conditions for these equations
are defined on the brane through Eq.~(\ref{brane}).  We emphasise once again
that, to obtain a complete braneworld theory in the general case
(including a stability analysis), one must also
specify additional conditions in the bulk such as the presence of other branes
or certain regularity conditions.  In our paper, we deal only with the
homogeneous and isotropic cosmology on the brane, so this issue does not arise.
In this sense, we are studying here the cosmological features common to the
whole class of braneworld models described by action (\ref{action}) with
arbitrary boundary conditions in the bulk.

\section{Vacuum branes and static branes} \label{vacua}

We proceed to the cosmological implications of the braneworld theory under
consideration. In this section, we discuss the situation pertaining to a vacuum
brane, i.e., when the matter stress-energy tensor $\tau_{ab} = 0$.  It is
interesting that the brane approaches this condition during the course of
cosmological evolution provided it expands forever and its matter density
asymptotically declines to zero. In this case, Eq.~(\ref{closed}) takes the
form
\begin{equation} \label{vacuum}
\left(M^6 + \frac23 \sigma m^2 \right) R + m^4 \left( R_{ab} R^{ab} - \frac13
R^2 \right) - 4 M^6 \Lambda_{\rm RS} = 0 \, ,
\end{equation}
where
\begin{equation} \label{lambda-rs}
\Lambda_{\rm RS} = {\Lambda_{\rm b} \over 2} + {\sigma^2 \over 3 M^6} \, .
\end{equation}
It is important to note that the second term in Eq.~(\ref{vacuum}) has {\em
precisely\/} the form of one of the terms in the expression for the conformal
anomaly, which describes the vacuum polarization at the one-loop level in
curved space-time (see, e.g., \cite{BD}).\footnote{It is interesting that,
while the conformal anomaly term $R_{ab} R^{ab} - \frac13R^2$
cannot be obtained by the
variation of a local four-dimensional Lagrangian, the very same
term is obtained via the variation of a
local Lagrangian in the five-dimensional braneworld theory under investigation.}
It therefore immediately follows that all {\em symmetric spaces\/} are
solutions of Eq.~(\ref{vacuum}) with appropriate $\Lambda_{\rm RS}$, just as
they are solutions of the Einstein equations with one-loop
quantum-gravitational corrections \cite{SK}. Symmetric spaces satisfy the
condition $\nabla_a R_{bcde} = 0$, which implies that geometrical invariants
such as $R_{abcd}R^{abcd}, R_{ab}R^{ab}$, and $R$ are constants so that
Eq.~(\ref{vacuum}) becomes an algebraic equation. Prominent members of this
family include:

(i) The homogeneous and isotropic de~Sitter space-time
\begin{equation} \label{desitter}
ds^2 = - dt^2 + \frac{1}{H^2} \cosh^2{Ht} \left[ d\chi^2 + \sin^2 \chi \left( d
\theta^2 + \sin^2 \theta d \phi^2 \right)\right] \, ,
\end{equation}
where $-\infty < t < \infty$, $0 \leq \chi,\theta \leq \pi$,
$0 \leq \phi \leq 2\pi$. The four dimensional metric (\ref{desitter}) has the
property $R^a{}_b = 3 H^2 h^a{}_b$, and formed the basis for
Starobinsky's first inflationary model sustained by the quantum conformal
anomaly \cite{star}.

(ii) The homogeneous and anisotropic Nariai metric \cite{nariai,kss}
\begin{equation}
ds^2 = k^2 \left( - dt^2 + \cosh^2{t} dr^2 + d\theta^2 + \sin^2\theta d\phi^2
\right) \, ,
\end{equation}
where $k$ = constant, $-\infty < t < \infty$, $0 \leq \theta \leq \pi$, $0 \leq
\phi \leq 2\pi$ and for which $R^a{}_b = h^a{}_b/k^2$. In fact, it is easy to
show that any metric for which $R$ and $R_{ab} R^{ab}$ are constants will
automatically be a solution to Eq.~(\ref{vacuum}) with an appropriate choice of
$\Lambda_{\rm RS}$.

Both de~Sitter space and the Nariai metric belong to the class of space-times
which satisfy the vacuum Einstein equations with a cosmological constant
\begin{equation}
R_{ab} = \Lambda h_{ab} \, .
\end{equation}
Such space-times also satisfy Eq.~(\ref{vacuum}) if
\begin{equation} \label{lambda}
\Lambda = {1 \over m^2} \left[ \left( {3 M^6 \over 2 m^2} + \sigma \right) \pm
\sqrt{ \left( {3 M^6 \over 2 m^2} + \sigma \right)^2 - 3 M^6 \Lambda_{\rm
RS}}\, \right] \, .
\end{equation}
Equation (\ref{lambda}) expresses the resulting cosmological constant on the
brane in terms of the coupling constants of the theory. For the frequently
discussed special case $m = 0$, one obtains $\Lambda = \Lambda_{\rm RS}$. The
two signs in (\ref{lambda}) correspond to the two different ways in which the
lower-dimensional brane can form the boundary of the higher-dimensional bulk
\cite{CH,Deffayet}. In the case of a spherically symmetrical bulk, the `$-$'
sign corresponds to a brane as a boundary for which the {\em inner\/} normal to
the brane points in the direction of decreasing bulk radial coordinate.

The condition $\Lambda_{\rm RS} = 0$ is the well-known fine-tuning condition of
Randall and Sundrum \cite{RS} and leads to the vanishing of the cosmological
constant on an empty brane if we set $m=0$ in (\ref{action}). Note that, under
the Randall--Sundrum condition, expression (\ref{lambda}) with the sign
opposite to the sign of the quantity $3 M^6 / 2 m^2 + \sigma$ also gives a zero
value for the resulting cosmological constant on the brane, but the other sign
usually leads to $\Lambda \neq 0$.

We would like to draw the reader's attention to the fact that
Eq.~(\ref{lambda}) is meaningful only when the expression under the square root
is nonnegative. When it is negative, solutions describing the corresponding
empty universe simply do not exist. This leads to the following important
conclusion: a universe which contains matter and satisfies
\begin{equation}
{3 M^6 \Lambda_{\rm RS} \over \left( 3 M^6 / 2 m^2 + \sigma \right)^2 }
> 1 \, ,
\end{equation}
{\em cannot expand forever}.

For the special case $3 M^6 / 2 m^2 + \sigma = 0$, the expression for $\Lambda$
is given by
\begin{equation}
\Lambda = \pm {M^3 \over m^2} \sqrt{- 3 \Lambda_{\rm RS}} \, .
\end{equation}
In this case, both $\sigma$ and $\Lambda_{\rm RS}$ must be negative in order
that the corresponding empty universe exist, but the resulting cosmological
constant on the brane can be of any sign.

Another interesting example is that of a static empty universe. The radius (scale
factor) $a$ of such a universe is easily determined from (\ref{vacuum}) to
be
\begin{equation} \label{static}
a^2 = {\kappa \over \Lambda_{\rm RS} } \left( \frac32  + {\sigma m^2 \over M^6}
\right)  \, ,
\end{equation}
where $\kappa = \pm 1$ is the sign of the spatial curvature.  One can see that
the radius of the universe can be arbitrarily large. In the general case, the
development of this solution to the five-dimensional bulk leads to a
Schwarzschild--anti-de~Sitter metric. It was shown in \cite{Shtanov1} that, for
$\kappa = 1$, this metric is purely anti-de~Sitter (with zero Schwarzschild
mass) if the constants of the theory satisfy the condition
\begin{equation}
{\sigma \over m^2} - {\Lambda_{\rm b} \over 2} + {3 M^6 \over 4 m^4} = 0 \, ,
\end{equation}
which implies negative brane tension $\sigma$. It should be pointed out that
the static and empty braneworld solution described by (\ref{static}) does not
possess a general-relativistic analog, since, in general relativity, a static
cosmological model (the `static Einstein universe') {\em cannot\/} be empty
(see, for instance, \cite{ss00}). Furthermore, from (\ref{static}) we find that
the static empty universe can be spatially open ($\kappa = -1$) --- for
example, in the case $\Lambda_{\rm RS} < 0$ and $\sigma > - 3 M^6 / 2 m^2$,
--- again a situation without an analog in general relativity.

For static homogeneous and isotropic braneworlds filled with matter,
Eq.~(\ref{closed}) gives the following relation:
\begin{equation}
a^2 \left[ \rho_{\rm tot} (\rho_{\rm tot} + 3 p_{\rm tot}) - 3 \Lambda_{\rm b}
M^6 \right] = 3 \kappa \left[ m^2 (\rho_{\rm tot} + 3 p_{\rm tot}) - 3 M^6
\right] \, ,
\end{equation}
where the total energy density $\rho_{\rm tot}$ and pressure $p_{\rm tot}$
include the contribution from the brane tension, i.e.,
\begin{equation}
\rho_{\rm tot} = \rho + \sigma \, , \quad p_{\rm tot} = p - \sigma \, ,
\end{equation}
and $\kappa = 0, \pm1$ corresponds to the sign of the spatial curvature.  This
relation reduces to (\ref{static}) for $\rho = p = 0$.

Having obtained all these solutions on the brane, one can find the
corresponding solutions in the bulk by integrating Eqs.~(\ref{s}) and
(\ref{metric}) with the initial conditions on the brane given by
Eq.~(\ref{brane}), as described in Sec.~\ref{basic}.  In doing this, one can
consider various additional conditions in the bulk, for example, existence of
other branes, or one can impose certain regularity conditions.  It is worth
noting that one and the same cosmological solution on the given brane can
correspond to different global solutions in the bulk, for example, other branes
may be present or absent, be static or evolving, close or far away from our
brane, etc.  In the most general case (for instance in the absence of special
symmetries on the brane) integration on the brane needs to be performed in
conjunction with dynamical integration in the bulk.
All such situations must be separately studied and issues such as
their stability to linearized perturbations must be examined on a
case-by-case basis.

\section{Cosmological Consequences of Braneworld Dark Energy} \label{cosmology}
\subsection{Cosmological evolution}

In the homogeneous and isotropic cosmological setting, Eq.~(\ref{closed}) can
be integrated with respect to time with the result \cite{Shtanov2}
\begin{equation} \label{cosmo}
m^4 \left( H^2 + {\kappa \over a^2} - {\rho + \sigma \over 3 m^2} \right)^2 =
M^6  \left(H^2 + {\kappa \over a^2} - {\Lambda_{\rm b} \over 6} - {C \over a^4}
\right) \, ,
\end{equation}
where  $C$ is the integration constant that corresponds to the black-hole mass
of the Schwarzschild--(anti)-de~Sitter solution in the bulk, $H \equiv \dot
a/a$ is the Hubble parameter on the brane, and $\kappa = 0, \pm 1$ corresponds
to the sign of the spatial curvature on the brane.  A generalization of this
equation to the case of absence of $Z_2$ symmetry was obtained in
\cite{KLM,Shtanov2}. For the frequently considered special case $m = 0$, this
equation reduces to the familiar one \cite{BDL}
\begin{equation}\label{cosmolim}
H^2 + {\kappa \over a^2} = {\Lambda_{\rm b} \over 6} + {C \over a^4} + {(\rho +
\sigma)^2 \over 9 M^6}\, .
\end{equation}

Equation (\ref{cosmo}) can be solved with respect to the Hubble parameter with
the result \cite{CH,Deffayet}:
\begin{equation} \label{solution}
H^2 + {\kappa \over a^2} = {\rho + \sigma \over 3 m^2} + {2 \over \ell^2}
\left[1 \pm \sqrt{1 + \ell^2 \left({\rho + \sigma \over 3 m^2} - {\Lambda_{\rm
b} \over 6} - {C \over a^4} \right)} \right] \, ,
\end{equation}
or, equivalently,
\begin{equation} \label{solution1}
H^2 + {\kappa \over a^2} = {\Lambda_{\rm b} \over 6} + {C \over a^4} + {1 \over
\ell^2} \left[\sqrt{1 + \ell^2 \left({\rho + \sigma \over 3 m^2} -
{\Lambda_{\rm b} \over 6} - {C \over a^4} \right)} \pm 1 \right]^2 \, ,
\end{equation}
where the length scale
\begin{equation}
\ell = {2 m^2 \over M^3}
\end{equation}
was introduced earlier in Sec.~\ref{basic}.  Again, the `$\pm$'signs in
(\ref{solution}) and (\ref{solution1}) correspond to two different ways of
bounding the Schwarzschild--(anti)-de~Sitter bulk space by the brane
\cite{CH,Deffayet}: the lower sign corresponds to the case where the inner
normal to the brane (which bounds the bulk) points in the direction of
decreasing bulk radial coordinate. Alternatively, the two different signs in
(\ref{solution}) and (\ref{solution1}) could correspond to the two possible
signs of the five-dimensional Planck mass $M$. Henceforth we shall refer to
models with the lower `$-$' sign as BRANE1 and models with the upper `$+$' sign
as BRANE2.

In the ensuing discussion, we shall neglect the decaying term $C / a^4$ in
Eq.~(\ref{solution}) (called dark radiation), assuming it to be negligibly
small at present.  Similarly to \cite{DDG}, we also introduce the dimensionless
cosmological parameters
\begin{equation} \label{omegas}
\Omega_{\rm m} =  {\rho_0 \over 3 m^2 H_0^2} \, , \quad \Omega_\kappa = -
{\kappa \over a_0^2 H_0^2} \, , \quad \Omega_\sigma = {\sigma \over 3 m^2
H_0^2} \, , \quad \Omega_\ell = {1 \over \ell^2 H_0^2} \, , \quad
\Omega_{\Lambda_{\rm b}} = - {\Lambda_{\rm b} \over 6 H_0^2} \, ,
\end{equation}
where the subscript `{\small 0}' refers to current values of cosmological
quantities. The system of cosmological equations with the energy density $\rho$
dominated by dust-like matter can now be written in the transparent form using
Eq.~(\ref{solution}):

\begin{itemize}
\item{BRANE1}
\begin{equation} \label{hubble1}
{H^2(z) \over H_0^2} = \Omega_{\rm m} (1\!+\!z)^3 + \Omega_\kappa (1\!+\!z)^2 +
\Omega_\sigma + \underline{2 \Omega_\ell - 2 \sqrt{\Omega_\ell}\,
\sqrt{\Omega_{\rm m} (1\!+\!z )^3 + \Omega_\sigma + \Omega_\ell +
\Omega_{\Lambda_{\rm b}}}} \, ,
\end{equation}
\item{BRANE2}
\begin{equation} \label{hubble2}
{H^2(z) \over H_0^2} = \Omega_{\rm m} (1\!+\!z)^3 + \Omega_\kappa (1\!+\!z)^2 +
\Omega_\sigma + \underline{2 \Omega_\ell + 2 \sqrt{\Omega_\ell}\,
\sqrt{\Omega_{\rm m} (1\!+\!z )^3 + \Omega_\sigma + \Omega_\ell +
\Omega_{\Lambda_{\rm b}}}} \, ,
\end{equation}
\end{itemize}
where $z$ is the cosmological redshift. We have underlined the terms which
cause our equations to differ from their general-relativistic counterparts.
The constraint equations for the cosmological parameters are most conveniently
derived using Eq.~(\ref{solution1}).  We obtain
\begin{equation} \label{original}
1 - \Omega_\kappa + \Omega_{\Lambda_{\rm b}} = \left( \sqrt{\Omega_\ell +
\Omega_{\rm m} + \Omega_\sigma + \Omega_{\Lambda_{\rm b}}} \pm
\sqrt{\Omega_\ell} \right)^2 \, .
\end{equation}
This equation can be transformed to a more convenient form. In doing this, we
note that the theory makes sense only in the case $\Omega_\ell + \Omega_{\rm m}
+ \Omega_\sigma + \Omega_{\Lambda_{\rm b}} > 0$. Since $\Omega_\ell > 0$, when
taking the square root of both sides of this equation, we must consider two
possibilities for the sign of the quantity $\Omega_{\rm m} + \Omega_\sigma +
\Omega_{\Lambda_{\rm b}}$ under the square root in (\ref{original}).

In the case
\begin{equation} \label{eq:first}
\Omega_{\rm m} + \Omega_\sigma + \Omega_{\Lambda_{\rm b}} \ge 0 \, ,
\end{equation}
taking the square root of both sides of (\ref{original}) and rearranging terms,
we get
\begin{equation}\label{eq:intermed1}
\sqrt{1 - \Omega_\kappa + \Omega_{\Lambda_{\rm b}}} \mp \sqrt{\Omega_\ell} =
\sqrt{\Omega_\ell + \Omega_{\rm m} + \Omega_\sigma + \Omega_{\Lambda_{\rm b}}}
\, .
\end{equation}
Now taking the square of both sides of the last equation, we finally obtain the
constraint:
\begin{equation} \label{con1}
\Omega_{\rm m} + \Omega_\kappa + \Omega_\sigma \pm 2 \sqrt{\Omega_\ell}\,
\sqrt{1 - \Omega_\kappa + \Omega_{\Lambda_{\rm b}}} = 1 \, .
\end{equation}

In the opposite case
\begin{equation} \label{eq:second}
- \Omega_\ell~ \leq ~\Omega_{\rm m} + \Omega_\sigma + \Omega_{\Lambda_{\rm b}}
< 0 \, ,
\end{equation}
taking the square root of both sides of (\ref{original}) and rearranging terms,
we get
\begin{equation}\label{eq:intermed2}
\sqrt{1 - \Omega_\kappa + \Omega_{\Lambda_{\rm b}}} - \sqrt{\Omega_\ell} = \pm
\sqrt{\Omega_\ell + \Omega_{\rm m} + \Omega_\sigma + \Omega_{\Lambda_{\rm b}}}
\, .
\end{equation}
Taking the square of both sides of the last equation, we finally obtain the
constraint:
\begin{equation} \label{strange}
\Omega_{\rm m} + \Omega_\kappa + \Omega_\sigma + 2 \sqrt{\Omega_\ell}\, \sqrt{1
- \Omega_\kappa + \Omega_{\Lambda_{\rm b}}} = 1 \, .
\end{equation}

\begin{figure}[tbh!]
\centerline{ \psfig{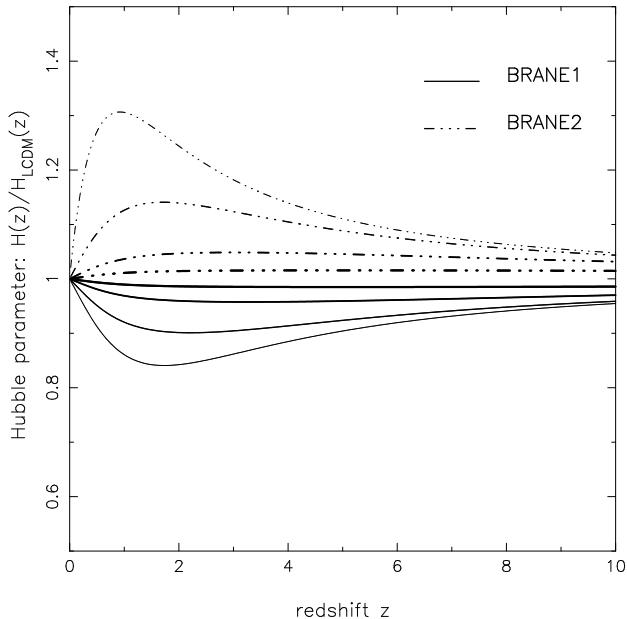} }
\bigskip
\caption{\small The Hubble parameter in units of $H_{\rm LCDM}(z)$ is plotted
as a function of redshift for the two braneworld models BRANE1 and BRANE2\@.
Whereas $H(z)$ in BRANE2 is {\em larger\/} than $H(z)$ in LCDM, in the case of
BRANE1 the value of $H(z)$ is {\em smaller\/} than its value in LCDM\@.
Parameter values are: $\Omega_\kappa = 0, ~\Omega_\ell = 1.0$, $\Omega_{\rm m}
= 0.3$, and $\Omega_{\Lambda_{\rm b}} = 1, 10, 10^2, 10^3$ (top to bottom for
BRANE2 and bottom to top for BRANE1). The value of $\Omega_\sigma$ is
determined from (\ref{omega-r1}) \& (\ref{omega-r2}). Parameter values for the
LCDM model are $\Omega_{\rm m} = 0.3, ~\Omega_\Lambda = 0.7$. For large values
of the bulk cosmological constant $\Omega_{\Lambda_{\rm b}}$, BRANE1 and BRANE2
become indistinguishable from LCDM. } \label{fig:hubble}
\end{figure}

To summarise, in the case $\Omega_{\rm m} + \Omega_\sigma +
\Omega_{\Lambda_{\rm b}} \ge 0$, the constraint equations in the BRANE1 and
BRANE2 models read:
\begin{equation} \label{omega-r1}
\Omega_{\rm m} + \Omega_\kappa + \Omega_\sigma - \underline{2
\sqrt{\Omega_\ell}\, \sqrt{1 - \Omega_\kappa + \Omega_{\Lambda_{\rm b}}}} = 1
\, ~~~({\rm BRANE1}) ,
\end{equation}
\begin{equation} \label{omega-r2}
\Omega_{\rm m} + \Omega_\kappa + \Omega_\sigma + \underline{2
\sqrt{\Omega_\ell}\, \sqrt{1 - \Omega_\kappa + \Omega_{\Lambda_{\rm b}}}} = 1
\, . ~~~({\rm BRANE2})
\end{equation}
In the opposite case $\Omega_{\rm m} + \Omega_\sigma + \Omega_{\Lambda_{\rm b}}
< 0$, both constraint equations for BRANE1 and BRANE2 are given by
Eq.~(\ref{omega-r2}).  Again, we have underlined the terms which cause our
equations to differ from their general-relativistic counterparts.  It is easy
to see that the general-relativistic limit of this theory is reached as
$\Omega_\ell \to 0$ (i.e., $M \to 0$). In this case, setting $\Omega_\kappa =
0, ~\Omega_\sigma \equiv \Omega_\Lambda$, we find that (\ref{hubble1}) \&
(\ref{hubble2}) give the expression for the Hubble parameter in the LCDM model:
\begin{equation}
\label{lcdm} H^2_{\rm LCDM}(z) = H_0^2\left\lbrack \Omega_{\rm m}(1 + z)^3 +
\Omega_\Lambda \right\rbrack.
\end{equation}
An example of the behaviour of the Hubble parameter in the braneworld models is
illustrated in Fig.~\ref{fig:hubble}.

It is convenient to rewrite Eq.~(\ref{lambda-rs}) in terms of the dimensionless
cosmological parameters
\begin{equation}
\Omega_{\rm RS} = \frac{\Lambda_{\rm RS}}{3H_0^2} = \frac{\Omega_\sigma^2} {4
\Omega_\ell} - \Omega_{\Lambda_{\rm b}} \, ,
\end{equation}
from where it follows that
the Randall--Sundrum constraint $\Lambda_{\rm RS} = 0$ has the simple form
\begin{equation} \label{RS}
\Omega_\sigma^2 - 4 \Omega_\ell \Omega_{\Lambda_{\rm b}} = 0 \, .
\end{equation}
This constraint implies $\Omega_{\Lambda_{\rm b}} \ge 0$ ($\Lambda_{\rm b} \leq
0$) and ensures the vanishing of the effective cosmological constant in the
future for the braneworld embedding described by the BRANE1 model for
$\Omega_\sigma \ge 0$ or for $\Omega_\sigma < 0$ and $\Omega_\ell >
\Omega_{\Lambda_{\rm b}}$, and by the BRANE2 model for $\Omega_\sigma < 0$ and
$\Omega_\ell < \Omega_{\Lambda_{\rm b}}$. The case where the Randall--Sundrum
constraint is imposed will be discussed in detail in Sec.~\ref{sec:RS} below.
It will be shown that, in the case of BRANE1 model, for the case $\Omega_\kappa
= 0$, Eq.~(\ref{omega-r1}) would imply that $\Omega_{\rm m} > 1$, which
apparently contradicts current observations of the large-scale distribution of
matter. This is an important result since it suggests that for reasonable
values of the matter density ($\Omega_{\rm m} \ll 1$) and $\Omega_\kappa = 0$,
the BRANE1 model demands the presence of a nonvanishing effective cosmological
constant. The value of the cosmological constant is given by Eq.~(\ref{lambda})
and can be determined by using observational data to fix the constants of the
theory.  On the other hand, the BRANE2 model is in principle observationally
compatible with the Randall--Sundrum constraint, which can result in vanishing
effective cosmological constant in the future thus making the current
acceleration of the universe a transient phenomenon.

Typical values of the $\Omega$ parameters (\ref{omegas}) that we consider in
this paper are of order $\Omega \sim 1$.  For such values, the fundamental
constants of the theory have the following orders of magnitude:
\begin{equation}
m^2 \simeq M_{\rm P}^2 \sim 10^{19}~\mbox{GeV}\, , \quad M \sim 100~\mbox{MeV}
\, , \quad \Lambda_{\rm b} \sim {\sigma \over m^2} \sim H_0^2 \sim 10^{-
56}~\mbox{cm}^{-2} \, .
\end{equation}
The smallness of the bare cosmological constants in the bulk and on the brane
represents a fine-tuning similar to what is the case for the cosmological
constant in the standard LCDM model.  However, even with such small values of
the bare cosmological constants in action (\ref{action}), the braneworld
evolution exhibits some qualitatively new properties (discussed in the
following subsections) when compared to the case where these constants are set
to zero. This is the reason why we find it important to keep open the
possibility of nonzero, albeit small, values of $\Lambda_{\rm b}$ and $\sigma$.

A possible procedure for testing the braneworld model against observations is
presented in the appendix.

\subsection{Cosmological tests of braneworld models}

Observations of high-redshift type Ia supernovae indicate that these objects
are fainter than they would be in a standard cold dark matter cosmology (SCDM)
with $\Omega_{\rm m} = 1$ \cite{sn}. This observation is taken as support for a
universe which is accelerating, fuelled by a form of energy with negative
pressure (dark energy).
In standard FRW cosmology the acceleration of the universe is described by
the equation
\beq\label{eq:acc}
\frac{\ddot a}{a} = -\frac{4\pi G}{3}\sum_i (\rho_i + 3p_i),
\eeq
where the summation is over all matter fields contributing to the
dynamics of the universe.
It is easy to show that a necessary (but not sufficient) condition for
acceleration (${\ddot a} > 0$) is that the strong energy condition
is violated for {\em at least one} of the matter fields in
(\ref{eq:acc}), so that $\rho + 3p < 0$.
In the case of the popular $\Lambda$CDM model this requirement is clearly
satisfied since $p_m = 0$ in pressureless (cold) matter, while
$p_\Lambda = -\rho_\Lambda \equiv -\Lambda/8\pi G$ in the cosmological
constant.
The situation with respect to braneworld models is different since, as we have demonstrated in the previous section, braneworld evolution is distinct
from FRW evolution at late times. However it is easy to show that braneworld
models can accelerate. We demonstrate this by noting that a completely general expression for the deceleration parameter $q = -{\ddot a}/aH^2$ is
provided by
\begin{equation} \label{decel}
q(z) = \frac{H'(z)}{H(z)} (1+z) - 1 \, ,
\end{equation}
where $H(z)$ is given by (\ref{hubble1})--(\ref{omega-r2})
and the derivative is with respect to $z$. The
current value of the deceleration parameter is easily calculated to be
\begin{equation}
q_0 = \frac32 \Omega_{\rm m} \left( 1 \pm \sqrt{\Omega_\ell \over \Omega_{\rm
m} + \Omega_\sigma + \Omega_\ell + \Omega_{\Lambda_{\rm b}}} \right) - 1 \, ,
\end{equation}
where the lower and upper signs correspond to BRANE1 and BRANE2 models,
respectively. Thus the present universe will accelerate ($q_0 < 0$) for
brane parameter values which satisfy
\beq
\frac32 \Omega_{\rm m} \left( 1 \pm \sqrt{\Omega_\ell \over \Omega_{\rm
m} + \Omega_\sigma + \Omega_\ell + \Omega_{\Lambda_{\rm b}}} \right) < 1.
\eeq

It should be pointed out that the proposition to use the induced curvature term
on a braneworld to account for the accelerated phase of our universe was first
made in \cite{Deffayet} in the context of the braneworld model of \cite{DGP},
which is the special case of our BRANE2 model with $\Omega_\sigma =
\Omega_{\Lambda_{\rm b}} = 0$. It was subsequently discussed and tested in
\cite{DDG,AM}. In the present paper, we allow both $\Omega_\sigma \neq 0$ and
$\Omega_{\Lambda_{\rm b}} \neq 0$, and the presence of these two free
parameters makes the braneworld model more flexible and allows for
qualitatively new behaviour which shall be discussed in detail in this section.
The connection between large extra dimensions and dark energy has also been
investigated in \cite{ABRS}.

Observationally, a pivotal role in the case for an accelerating universe
is played by the {\em
luminosity distance\/} $d_L(z)$, since the flux of light received from a
distant source varies inversely to the square of the luminosity distance,
${\cal F} \propto d_L^{-2}$. This effect is quantitatively described by the
magnitude--luminosity relation: $m_B = M_0 + 25 + 5\log_{10}{d_L}$, where $m_B$
is the corrected apparent peak $B$ magnitude and $M_0$ is the absolute peak
luminosity of the supernova. A supernova will therefore appear fainter in a
universe which possesses a larger value of the luminosity distance to a given
redshift.

In a FRW universe, the luminosity distance is determined by the Hubble
parameter and three-dimensional spatial curvature \cite{ss00}:
\begin{equation}
d_L(z) = {1 + z \over H_0 \sqrt{\left| \Omega_{\rm total} - 1 \right|}}\, S
\left( \eta_0 - \eta \right) \, ,  \label{eq:age11d}
\end{equation}
where
\begin{equation}
\eta_0 - \eta = H_0 \sqrt{ \left| \Omega_{\rm total} - 1
\right|} \int_0^z {dz' \over H(z')} \, , \label{lumdis1}
\end{equation}
and
$S(x)$ is defined as follows: $S(x) = \sin x$ if $\kappa = 1$ ($\Omega_{\rm
total} > 1$), $S(x) = \sinh x$ if $\kappa = -1$ ($\Omega_{\rm total} < 1$), and
$S(x) = x$ if $\kappa = 0$ ($\Omega_{\rm total} = 1$).

\begin{figure}[tbh!]
\centerline{ \psfig{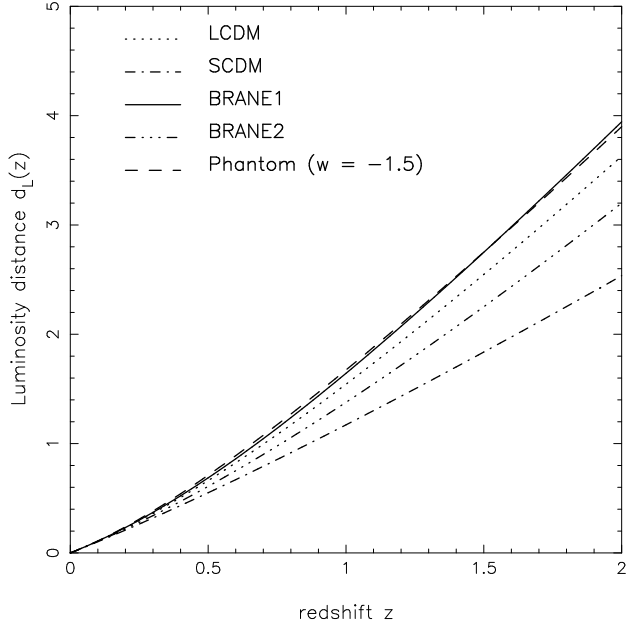} }
\bigskip
\caption{\small The luminosity distance is shown as a function of redshift for
the two braneworld models BRANE1 \& BRANE2, LCDM, SCDM, and `phantom energy'.
All models, with the exception of SCDM, have $\Omega_{\rm m} = 0.3$. SCDM has
$\Omega_{\rm m} = 1$. The BRANE1 \& BRANE2 models have $\Omega_\ell = 0.3$ and
vanishing cosmological constant in the bulk. LCDM and the phantom model have
the same dark energy density $\Omega_\Lambda = \Omega_X = 0.7$. The equation of
state for dark energy is $w_\Lambda = -1$ for LCDM and $w = p_X/\rho_X = -1.5$
for phantom. The luminosity distance is greatest for BRANE1 \& phantom, and
least for SCDM\@. BRANE1 \& BRANE2 lie on either side of LCDM. }
\label{fig:lum}
\end{figure}

Inflationary models suggest that the universe is nearly flat with
$\Omega_\kappa \simeq 0$ and we shall work under this assumption in the rest of
this paper (see also \cite{cmb}). In this case, Eq.~(\ref{eq:age11d})
simplifies to \begin{equation} {d_L(z) \over 1 + z} = \int_0^z {dz' \over
H(z')} \, , \label{lumdis2} \end{equation} where $H(z)$ is determined by
(\ref{hubble1}) for BRANE1, and by (\ref{hubble2}) for BRANE2\@. In
Fig.~\ref{fig:lum} we show the luminosity distances for the BRANE1 \& BRANE2
models. Also shown for comparison is the value of $d_L(z)$ in a spatially-flat
two-component FRW universe with the Hubble parameter \begin{equation} H(z) =
H_0\left[ \Omega_{\rm m} (1 + z)^3 + \Omega_X (1 + z)^{3(1 + w)} \right]^{1/2},
\label{hubble3} \end{equation} where $\Omega_X$ describes dark energy with
equation of state $w = p_X/\rho_X$. Three cosmological models will be of
interest to us in connection with (\ref{hubble3}):

\smallskip
(i) SCDM: The standard cold dark matter universe with $\Omega_{\rm m} = 1$ and
$\Omega_X = 0$.

\smallskip
(ii) LCDM: Cold dark matter $+$ a cosmological constant with $w = -1$.

\smallskip
(iii) Phantom models: Cold dark matter $+$ `phantom energy' satisfying $w < -1$
\cite{caldwell}.

\smallskip

We find from Fig.~\ref{fig:lum} that the luminosity distance in both braneworld
models exceeds that in SCDM\@. In fact, BRANE1 models have the unusual feature
that their luminosity distance can even exceed that in LCDM (for a fixed value
of $\Omega_{\rm m}$)\,\@! In fact it can easily be shown that \begin{equation}
d_L^{~\rm dS}(z) \geq d_L^{~\rm BRANE1}(z) \geq d_L^{~\rm LCDM}(z) \, ,
\end{equation} where $d_L^{~\rm dS}(z)$ refers to the luminosity distance in
the spatially flat coordinatization of de~Sitter space (equivalently, the
steady state universe). The second inequality presumes a fixed value of
$\Omega_{\rm m}$. BRANE2 models show complementary behaviour \begin{equation}
d_L^{~\rm LCDM}(z) \geq d_L^{~\rm BRANE2}(z) \geq d_L^{~\rm SCDM}(z) \, ,
\end{equation} where the first inequality is valid for a fixed value of
$\Omega_{\rm m}$.  In the case $\Omega_\ell = 0$, the equations of the
braneworld theory formally reduce to those of general relativity, and we have
$d_L^{~\rm BRANE1}(z) = d_L^{~\rm BRANE2}(z) = d_L^{~\rm LCDM}(z)$.

One might add that the behaviour of BRANE1 is mimicked by FRW models with $w
\leq -1$, whereas BRANE2 resembles dark energy with $-1 \leq w \leq 0$
\cite{ss00}. In fact, from Fig.~\ref{fig:lum} we see that the luminosity
distance in the BRANE1 model is quite close to what one gets from `phantom
energy' described by (\ref{hubble3}) with $w = -1.5$. (The parameters for this
BRANE1 model are $\Omega_{\rm m} = \Omega_\ell = 0.3$, $\Omega_{\Lambda_{\rm
b}} = 0$, and $\Omega_\sigma = 1 - \Omega_{\rm m} + 2 \sqrt{\Omega_\ell}
\approx 1.8$.)  It should be pointed out that phantom energy models were
introduced by Caldwell \cite{caldwell}, who made the important observation that
dark-energy with $w < -1$ appears to give a better fit to supernova
observations than LCDM (which has $w = -1$). Unfortunately, phantom models have
several bizarre properties, which may be the reason why cosmologists have been
reluctant to take these models seriously despite their seemingly better
agreement with observations (see however \cite{maor}). Some strange properties
of phantom energy with a very negative equation of state ($w < -1$) are
summarised below (see also \cite{caldwell,innes}):

(i) A negative equation of state suggests that the effective velocity
of sound in the medium $v = \sqrt{|dp/d\rho|}$ can become larger than
the velocity of light.

(ii) The expansion factor of a universe dominated by phantom energy grows as
\begin{equation}
a(t) \simeq a\left( t_{\rm eq} \right) \left[(1 + w) \frac{t}{t_{\rm eq}} - w
\right]^{2 \big/ 3(1+w)} \, , ~~ w < -1 ~,
\end{equation}
where $t_{\rm eq}$ marks the epoch when the densities in matter and phantom
energy are equal: $\rho_m(t_{\rm eq}) \simeq \rho_X(t_{\rm eq})$. It
immediately follows that the scale factor diverges in a {\em finite\/} amount
of cosmic time
\begin{equation}
a(t) \to \infty \ \ \mbox{as} \ \ t \to t_* = \left( \frac{w}{1+w} \right)
t_{\rm eq} \, .
\end{equation}
Substitution of $z \to -1$ and $w < -1$ in (\ref{hubble3}) shows that the
Hubble parameter also diverges as $t \to t_*$, implying that an infinitely
rapid expansion rate for the universe has been reached in the {\em finite\/}
future.

As the universe expands, the density of phantom energy ($w < -1$) {\em grows\/}
instead of decreasing ($w > -1$) or remaining constant ($w = -1$),
\begin{equation} \rho(t) \propto \left[(1 + w) \frac{t}{t_{\rm eq}} - w
\right]^{-2} , \end{equation} reaching a singular value in a finite interval of
time $\rho(t) \to \infty$, $t \to t_*$. This behaviour should be contrasted
with the density of ordinary matter which drops to zero: $\rho_m \to 0$ as $t
\to t_*$. A universe dominated by phantom energy is thus doomed to {\em expand
towards a physical singularity\/} which is reached in a finite amount of proper
time. (An exact expression for the time of occupance of the phantom singularity
is given in \cite{star99}, which also contains an interesting discussion of
related issues.)

\begin{figure}[tbh!]
\centerline{ \psfig{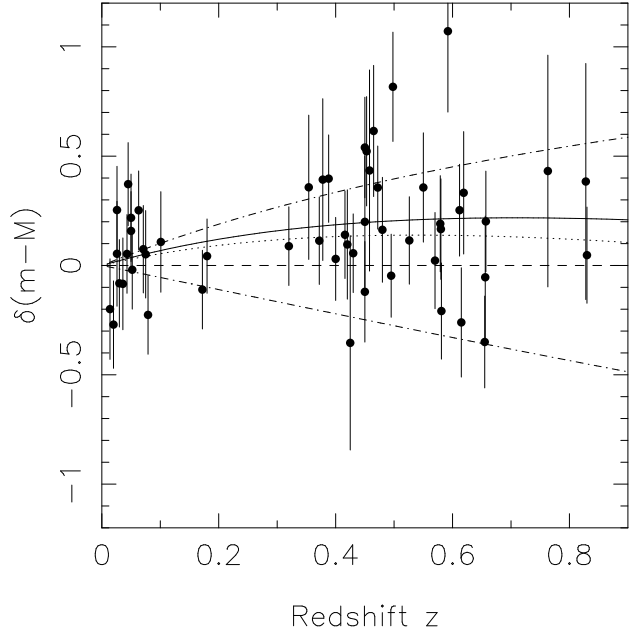} }
\bigskip
\caption{\small The distance modulus ($m-M$) of Type Ia supernovae (the primary
fit of the Supernova cosmology project) is shown relative to an empty
$\Omega_{\rm m} \to 0$ Milne universe (dashed line). The solid line refers to
the distance modulus in BRANE1 with $\Omega_\ell = \Omega_{\rm m} = 0.3$, and
vanishing cosmological constant in the bulk. The dotted line (below the solid)
is LCDM with $(\Omega_\Lambda, \Omega_{\rm m}) = (0.7,0.3)$. The uppermost and
lowermost (dot-dashed) lines correspond to de~Sitter space $(\Omega_\Lambda,
\Omega_{\rm m}) = (1,0)$ and SCDM $(\Omega_\Lambda, \Omega_{\rm m}) = (0,1)$,
respectively.} \label{fig:sn}
\end{figure}

At this stage one must emphasize that, although the BRANE1 model has several
features in common with phantom energy, (which makes us believe that, like the
latter, it too is likely to provide a good fit to supernova data), it is not
necessarily afflicted with phantom's pathologies. Indeed, in a broad range of
parameters, both BRANE1 and BRANE2 are physically well motivated and remain
{\em well behaved during all times\/}. The present paper therefore adds an
important new dimension to the current debate about the acceleration of the
universe by showing that cosmological models with $d_L(z) > d_L^{~\rm LCDM}(z)$
are possible to construct within the framework of the braneworld scenario and
should be taken seriously. Future studies will address the important
quantitative issue of whether braneworld models of dark energy provide as good
(or better) fit to high-$z$ supernova observations as LCDM (see
Fig.~\ref{fig:sn} as an illustration). (It is appropriate to mention here that
the weak energy condition $\rho + p \geq 0$ can also be violated in a class of
scalar-tensor theories of gravity, as discussed in \cite{beps}.)

\begin{figure}[tbh!]
\centerline{ \psfig{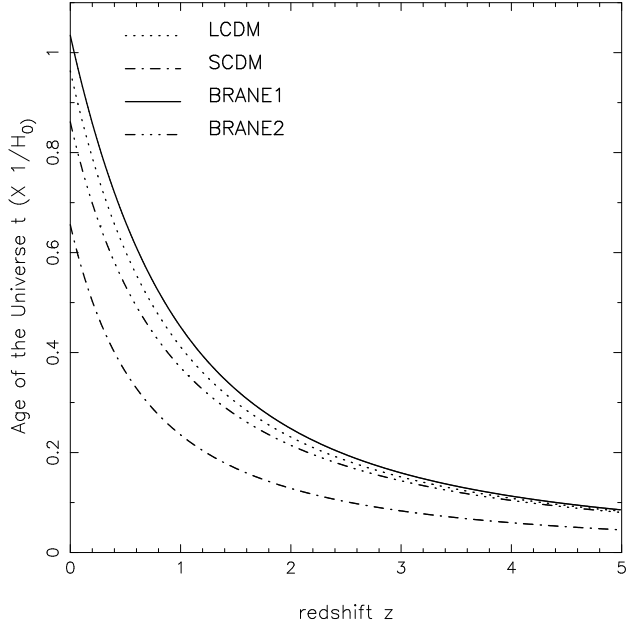} }
\bigskip
\caption{\small The age of the universe (in units of the inverse Hubble
parameter) is plotted as a function of the cosmological redshift for the models
discussed in Fig.~\ref{fig:lum}\@. (The phantom model is not shown.) BRANE1
models have the oldest age while SCDM is youngest. } \label{fig:age}
\end{figure}

\begin{figure}[tbh!]
\centerline{ \psfig{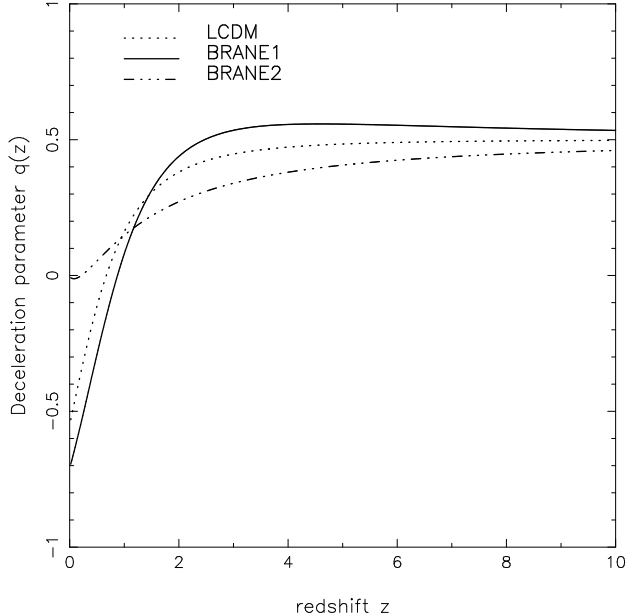} }
\bigskip
\caption{\small The deceleration parameter $q(z)$ is shown for BRANE1, BRANE2
and LCDM\@. The model parameters are as in Fig.~\ref{fig:lum}\@. For reference
it should be noted that $q = 0.5$ for SCDM while de~Sitter space has $q = -1$.
} \label{fig:decel}
\end{figure}

\begin{figure}[tbh!]
\centerline{ \psfig{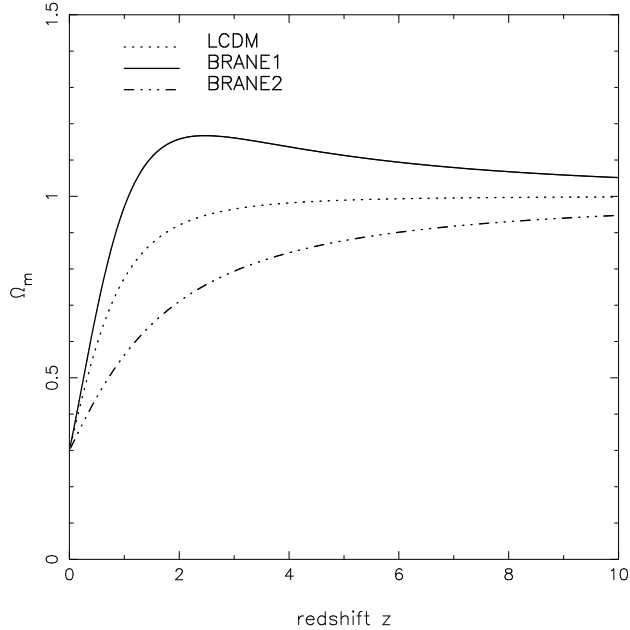} }
\bigskip
\caption{\small The dimensionless matter density $\Omega_{\rm m}(z)$ is shown
for the two braneworld models and LCDM\@. ($\Omega_{\rm m} = 1$ in SCDM\@.)
Parameter values are the same as in previous figures. BRANE1 has the
interesting feature that $\Omega_{\rm m}(z)$ {\em exceeds unity\/} for $z \gsim
1$. } \label{fig:omega}
\end{figure}

We should add that the `angular-size distance' $d_A$ is related to the
luminosity distance $d_L$ through $d_A = d_L (1+z)^{-2}$, therefore much of the
above analysis carries over when one discusses properties of the angular-size
distance within the framework of braneworld models. Some cosmological features
of braneworld models are shown in Figs.~\ref{fig:age}--\ref{fig:omega}. In
Fig.~\ref{fig:age}, the age of the universe at a given cosmological redshift
\begin{equation} t(z) = \int_z^\infty \frac{dz'}{(1+z') H(z')} \, , \end{equation} is shown for the
two braneworld models and for LCDM \& SCDM\@. We find that the age of the
universe in BRANE1 (BRANE2) is larger (smaller) than in LCDM for identical
values of the cosmological density parameter $\Omega_{\rm m}$. This is a direct
consequence of the fact that the Hubble parameter in BRANE1 (BRANE2) is smaller
(larger) than in LCDM\@. Both braneworld models are significantly older than
SCDM.

Considerable insight into the dynamics of the universe is provided by
the cosmological deceleration parameter (\ref{decel}).
Our results, shown in Fig.~\ref{fig:decel}, indicate that at late
times the BRANE1 (BRANE2) universe accelerates at a faster (slower) rate than
LCDM (with identical $\Omega_{\rm m}$). Curiously, the BRANE1 universe shows an
earlier transition from deceleration to acceleration than any of the other
models. (For the given choice of parameters this transition takes place at $z
\simeq 1$ for BRANE1 and at $z \simeq 0.7$ for LCDM\@. The BRANE2 model begins
accelerating near the present epoch at $z \simeq 0$.) A related point of
interest is that at $z \gsim 2$ the deceleration parameter in BRANE1 marginally
exceeds that in SCDM indicating that the BRANE1 model is decelerating at a
faster rate than SCDM ($q = 0.5$). In conventional models of dark matter this
behaviour can occur only if the equation of state of the dark component is
stiffer than dust, implying $w > 0$ in (\ref{hubble3}),
or if the universe is spatially closed. On the other hand, the
current {\em acceleration\/} rate of BRANE1 in our example ($q_0 \simeq -0.7$)
significantly exceeds that of LCDM ($q_0 \simeq -0.55$) with an identical value
of $\Omega_{\rm m} = 0.3$ in both models. Within the framework of
four-dimensional Einstein gravity, this situation can only arise if the
equation of state of dark energy is strongly negative: $w < -1$ in
(\ref{hubble3}).

The unusual high-$z$ behaviour of the deceleration parameter in BRANE1 can be
better understood if we consider the cosmological density parameter
\begin{equation}
\Omega_{\rm m}(z) = \left[ \frac{H_0}{H(z)} \right]^2 \Omega_{\rm m}(0) (1+z)^3
\, , \label{eq:omega}
\end{equation}
where $H(z)$ is given by (\ref{hubble1})--(\ref{omega-r2}) for braneworld
models. From Fig.~\ref{fig:omega} we notice that, for $z \gsim 1$,
 the value of $\Omega_{\rm m}(z)$
in BRANE1 {\em exceeds\/} its value in SCDM ($\Omega_{\rm m} = 1$). This is
precisely the redshift range during which $q(z)_{\rm BRANE1} > q(z)_{\rm
SCDM}$. Thus, the rapid deceleration of BRANE1 at high redshifts can be partly
attributed to the larger value of the matter density $\Omega_{\rm m}(z)$ at
those redshifts, relative to SCDM.

Having established partial similarity of BRANE1 with phantom models at low
redshifts, we can
investigate the analogy further and calculate the effective equation of state
of dark energy
\begin{equation} \label{eq:dark}
w(z) = {2 q(z) - 1 \over 3 \left[ 1 - \Omega_{\rm m}(z) \right] } \, ,
\end{equation}
where $\Omega_{\rm m}(z)$ is given by (\ref{eq:omega}). One notes that $w(z)$
has a pole-like singularity at $z \simeq 1$ for BRANE1, which arises because
$\Omega_{\rm m} (z)$ crosses the value of unity at $z \simeq 1$ (see
Fig.~\ref{fig:omega}). This demonstrates that the notion of `effective equation
of state' is of limited utility for this model. Equations (\ref{hubble1}),
(\ref{hubble2}), (\ref{decel}), and (\ref{eq:dark}) also illustrate the
important fact that dark energy in braneworld models, though similar to phantom
energy in some respects, differ from it in others. For instance, in both
braneworld models, $w(z) \to -0.5$ at $z \gg 1$ and $w(z) \to -1$ as $z \to
-1$, whereas phantom energy has $w(z) < -1$ at {\em all\/} times.

A useful quantity is the {\em current value\/} of the effective equation of
state of dark energy in braneworld theories:
\begin{equation}
w_0 = {2 q_0 - 1 \over 3 \left( 1 - \Omega_{\rm m} \right)} = - 1 \pm
{\Omega_{\rm m} \over 1 - \Omega_{\rm m}} \, {\sqrt{\Omega_\ell \over
\Omega_{\rm m} + \Omega_\sigma + \Omega_\ell + \Omega_{\Lambda_{\rm b}}}} \, ,
\end{equation}
where the lower and upper signs, as usual, correspond to BRANE1 and BRANE2
models, respectively.  We easily see that $w_0 < - 1$ for BRANE1, whereas
 $w_0 > - 1$ for BRANE2.

\subsection{Disappearing Dark Energy} \label{sec:RS}

The braneworld models under consideration also admit the intriguing possibility
that the current acceleration of the universe may not be a lasting feature. It
may be recalled that most models of dark energy including the cosmological
constant have the property that, once the universe begins to accelerate, it
accelerates forever. As shown in a number of recent papers, an eternally
accelerating universe is endowed with a cosmological event horizon which
prevents the construction of a conventional S-matrix describing particle
interactions within the framework of string or M-theory \cite{horizon}. In this
section we show that, provided the Randall--Sundrum constraint relation
(\ref{RS}) is satisfied, the acceleration of the universe can be a transient
phenomenon in braneworld models.  An anisotropic solution of Bianchi~V class
with the same feature was described in \cite{Kofinas}.

\begin{figure}[tbh!]
\centerline{ \psfig{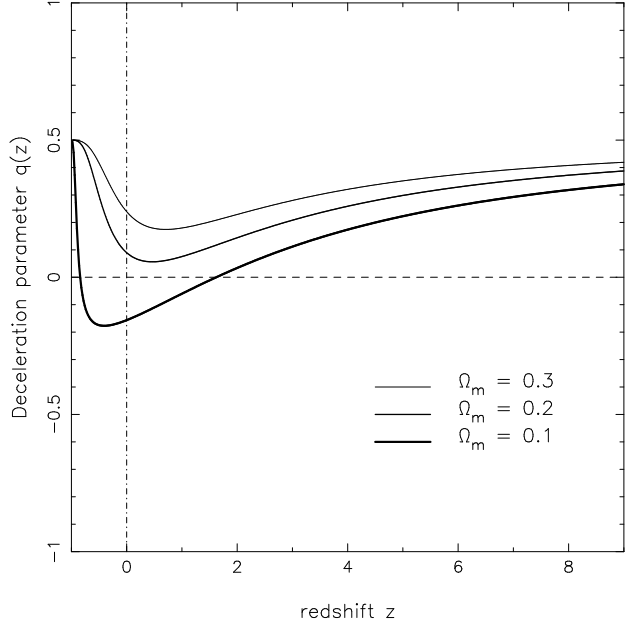} }
\bigskip
\caption{\small The deceleration parameter is plotted as a function of redshift
for the BRANE2 model with the Randall--Sundrum constraint (\ref{RS}),
$\Omega_{\Lambda_{\rm b}} = 2$, and $\left(\Omega_{\rm m}, \Omega_\ell \right)
= (0.3, 1.2)$, $(0.2, 1.6)$, $(0.1, 1.98)$ (top to bottom). The vertical
(dot-dashed) line at $z = 0$ marks the present epoch, while the horizontal
(dashed) line at $q = 0$ corresponds to a Milne universe [$a(t) \propto t$]
which neither accelerates nor decelerates. Note that the universe {\em ceases
to accelerate\/} and becomes matter dominated in the future.}
\label{fig:decel_plot}
\end{figure}

\begin{figure}[tbh!]
\centerline{ \psfig{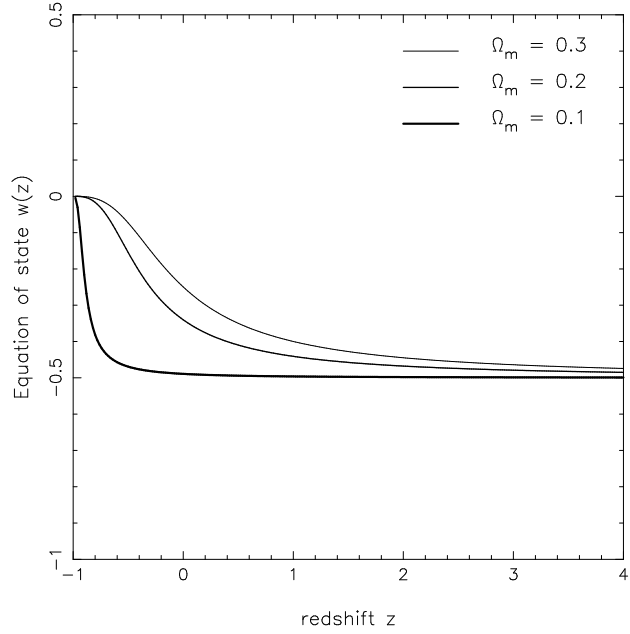} }
\bigskip
\caption{\small The effective equation of state for dark energy in the BRANE2
model is shown as a function of redshift. Model parameters are as in the
previous figure. Note that the past and future asymptotes of $w(z)$ are quite
different: $w(z) \to -1/2$ for $z \gg 0$, while $w(z) \to 0$ for $z \to -1$.
Braneworld dark energy therefore effectively disappears in the future, giving
rise to a matter-dominated universe. } \label{fig:state_future}
\end{figure}

From Eqs.~(\ref{hubble1}) and (\ref{hubble2}) we obtain the following
asymptotic expressions for the Hubble parameter $H_\infty$ as $z \to -1$,
assuming that the universe expands forever:
\begin{equation}
\left({H_\infty \over H_0}\right)^2 = \Omega_\sigma + 2 \Omega_\ell \pm 2
\sqrt{\Omega_\ell} \, \sqrt{\Omega_\sigma + \Omega_\ell + \Omega_{\Lambda_{\rm
b}}} \, ,
\end{equation}
where the lower and upper signs correspond to BRANE1 and BRANE2 models,
respectively.  In applying the Randall--Sundrum constraint (\ref{RS}), we first
consider the case where $\Omega_\sigma > 0$. Then
\begin{equation}
\Omega_\sigma = 2 \sqrt{\Omega_\ell \Omega_{\Lambda_{\rm b}}}
\end{equation}
and
\begin{equation} \label{positive}
\left({H_\infty \over H_0}\right)^2 = 2 \sqrt{\Omega_\ell} \left[
\sqrt{\Omega_\ell} + \sqrt{\Omega_{\Lambda_{\rm b}}} \pm \left(
\sqrt{\Omega_\ell} + \sqrt{\Omega_{\Lambda_{\rm b}}} \right) \right] \, .
\end{equation}
One can see that this expression vanishes for the lower sign.  Thus, for
positive $\Omega_\sigma$, it is the BRANE1 model that leads to vanishing
effective cosmological constant in the future.  However, in this case, the
constraint equation (\ref{omega-r1}) with $\Omega_\kappa = 0$ becomes
\begin{equation}
\Omega_{\rm m} - 2 \sqrt{\Omega_\ell} \left( \sqrt{1 + \Omega_{\Lambda_{\rm
b}}} - \sqrt{\Omega_{\Lambda_{\rm b}}} \right) = 1
\end{equation}
and implies $\Omega_{\rm m} > 1$, which is hardly compatible with the
observations.

In the case of $\Omega_\sigma < 0$, we have
\begin{equation} \label{eq:omsigma}
\Omega_\sigma = - 2 \sqrt{\Omega_\ell\, \Omega_{\Lambda_{\rm b}}}
\end{equation}
and
\begin{equation} \label{negative}
\left({H_\infty \over H_0}\right)^2 = 2 \sqrt{\Omega_\ell} \left(
\sqrt{\Omega_\ell} - \sqrt{\Omega_{\Lambda_{\rm b}}} \pm
\left|\sqrt{\Omega_\ell} - \sqrt{\Omega_{\Lambda_{\rm b}}} \right|\, \right) \,
.
\end{equation}
If $\Omega_\ell > \Omega_{\Lambda_{\rm b}}$, then this expression vanishes for
the lower sign, which brings us back to the nonphysical BRANE1 models with
$\Omega_{\rm m} > 1$. The case $\Omega_\ell > 1 + \Omega_{\Lambda_{\rm b}}$ is
compatible with the condition $\Omega_{\rm m} + \Omega_\sigma +
\Omega_{\Lambda_{\rm b}} < 0$, which corresponds to BRANE1 (\ref{hubble1}) with
the constraint equation (\ref{omega-r2}). However, it can be shown that these
conditions imply $\Omega_{\rm m} < 0$, which is also physically unacceptable.

There remains the case of $\Omega_\sigma < 0$ and $\Omega_\ell \le
\Omega_{\Lambda_{\rm b}}$.  In this case, expression (\ref{negative}) vanishes
for the upper sign, which corresponds to BRANE2 models. The constraint equation
(\ref{omega-r2}) with $\Omega_\kappa = 0$ now reads
\begin{equation} \label{eq:con-rs}
\Omega_{\rm m} + 2 \sqrt{\Omega_\ell} \left( \sqrt{1 + \Omega_{\Lambda_{\rm
b}}} - \sqrt{\Omega_{\Lambda_{\rm b}}} \right) = 1
\end{equation}
and implies $\Omega_{\rm m} < 1$.

Therefore, BRANE2 with $\Omega_\sigma < 0$ and $\Omega_\ell \le
\Omega_{\Lambda_{\rm b}}$, provides us with an interesting example of a
physically meaningful cosmological model in which the current acceleration of
the universe is a {\em transient phenomenon\/}. An example of this behaviour as
probed by the deceleration parameter is shown in Fig.~\ref{fig:decel_plot},
which demonstrates that the current period of cosmic acceleration takes place
between two matter-dominated epochs. We emphasize that these models require
negative brane tension $\sigma$.  Since an observer in this model resides on a
negative-tension brane one must ponder over the issue of whether such a
braneworld will be perturbatively stable and hence physically viable. We
consider this to be an open question for future investigations. Remarks made at
the end of Sec.~\ref{vacua} are relevant, however, since one and the same
cosmological solution on the `visible' (negative tension) brane can correspond
to many different global conditions in the bulk, for instance, other (`hidden')
branes may be present or absent, static or evolving, close to or far away from
our brane, etc.  Moreover, here the sign of the parameter $\epsilon = \pm 1$ in
action (\ref{action}) is expected to be of crucial importance, since, in
particular, the bulk gravity will be localised around the {\em negative\/}
tension (visible) brane in the case of $\epsilon = - 1$.  Issues of
perturbative stability involve the examination of all such distinct global
solutions on a case-by-case basis. Such a study, though interesting and
important, clearly lies beyond the scope of the present paper.

Useful insight into the BRANE2 model is also provided by the effective equation
of state of dark energy (\ref{eq:dark}). Our results shown in
Fig.~\ref{fig:state_future} indicate that the past and future behaviour of dark
energy in the braneworld universe can be very different. The past behaviour
$w(z) \to -0.5$ for $z \gg 1$ arises because, in a spatially flat braneworld,
the second most important contribution to braneworld expansion at high
redshifts is caused by the $(1+z)^{3/2}$ term in (\ref{hubble2}); see also
\cite{DDG}. The future behaviour $w(z) \to 0$ as $z \to -1$, on the other hand,
reflects the decreasing importance of dark energy as the universe expands. The
acceleration of the universe is therefore a transient phenomenon which ends
once the universe settles back into the matter-dominated regime.

Finally, we should mention that a transiently accelerating regime also arises
in a class of BRANE2 models which do not satisfy the Randall-Sundrum constraint
(\ref{RS}). In these models the current epoch of acceleration is succeeded by
an epoch during which the deceleration parameter grows without bound. This
unusual `future singularity' is reached in a {\em finite\/} interval of
expansion time and is characterised by the fact that both the matter density
and the Hubble parameter remain finite, while ${\ddot a} \to \infty$ (a feature
that distinguishes it from the phantom singularities discussed in Sec. IV B). A
detailed discussion of the `new' singularities which occur in braneworld models
can be found in \cite{ss02}.

\section{Conclusions} \label{final}

In this paper, we have considered cosmological implications of a braneworld
model described by action (\ref{action}), which contains both bulk and brane
curvature terms and cosmological constants. The curvature term for the brane
arises naturally, as a quantum correction from the matter part of the brane
action, and significantly changes the behaviour of the braneworld theory. For
example, as is well known, braneworld  cosmology without this term deviates
from general relativity at {\em large\/} matter densities, i.e., at early
cosmological times; see Eq.~(\ref{cosmolim}). In the model that we considered,
on the other hand, early cosmological evolution remains virtually unaffected
for a broad range of parameters, significant deviations from standard cosmology
appearing only during {\em later\/} times.

We restrict our attention to the important case where the brane forms the
boundary of the five-dimensional bulk space. This is equivalent to endowing the
bulk with the $Z_2$ reflection isometry with respect to the brane. The presence
of the curvature term in the action leads to two families of braneworld models.
These two families (called BRANE1 \& BRANE2) differ in the manner in which the
brane forms the boundary of the five-dimensional bulk. For example, if a
spatially three-spherical brane is embedded into the three-spherically
symmetric bulk, then one can discard either the interior part of the bulk, or
its exterior (asymptotically flat) part.  In this example, the models which
discard the exterior part leaving the interior were classified as BRANE1
models, while the complementary models were called BRANE2\@.  Similar
classification arises in the case of spatially flat and open braneworld models.
Alternatively, the two different families of braneworld models can be regarded
as corresponding to the two possible signs of the five-dimensional Planck mass
$M$.

Braneworld models of dark energy considered in this paper have the interesting
and unusual property that their luminosity distance $d_L$ can {\em exceed\/}
that in LCDM. This is unusual since, within the general-relativistic framework,
the luminosity distance has this property {\em only if\/} the equation of state
of dark energy is strongly negative ($w < -1$). Caldwell \cite{caldwell} has
shown that dark energy with $w < -1$ (phantom energy) may in fact fit supernova
observations better than LCDM\@. However, phantom energy is beset with a host
of undesirable properties which makes this model of dark energy unattractive.
We show that braneworld models have all the advantages and none of the
disadvantages of phantom models and therefore endow dark energy with exciting
new possibilities.
A recent analysis of braneworld models in \cite{alam} has demonstrated that
BRANE1 models (which generically have $w \leq -1$) provide an excellent fit to supernovae
observations for higher values of the matter density ($\Omega_m \ggeq 0.3$),
whereas lower values ($\Omega_m \lleq 0.25$) are preferred by BRANE2 models
(which generically have $w \geq -1$).

A distinctive feature of the braneworld scenario discussed in this paper is
that it allows for a universe which is {\em transiently accelerating}. Recent
investigations indicate that an eternally accelerating universe, which
possesses a cosmological event horizon, prevents the construction of a
conventional S-matrix describing particle interactions within the framework of
string or M-theory \cite{horizon}.  We have demonstrated that braneworld models
can enter into a regime of accelerated expansion at late times {\em even if\/}
the brane tension and the bulk cosmological constant are tuned to satisfy the
Randall--Sundrum constraint on the brane.  In this case, braneworld dark energy
and the acceleration of the universe are {\em transient\/} phenomena. In this
class of models, the universe, after the current period of acceleration,
re-enters the matter-dominated regime.  We have shown that viable models
realising this behaviour are those of BRANE2 type. (Since these braneworlds
have a negative brane tension the question of their stability is important and
needs to be examined in detail. We shall return to this issue in a companion
paper.) Thus braneworld models can give rise to a {\em transiently accelerating
phase} thereby reconciling a dark energy dominated universe with the
requirements of string/M-theory.  (A similar observation was previously made in
\cite{Kofinas} in the context of an anisotropic cosmological solution of
Bianchi~V class.)

\section*{Acknowledgments}
The authors acknowledge support from the Indo-Ukrainian program of cooperation
in science and technology. We also thank Tarun Deep Saini for several useful
discussions and for producing Figure~\ref{fig:sn}.

\appendix
\section*{Testing the model}

In this appendix, we present a possible procedure for testing the braneworld
model against observations.

There are the following four possibilities to be considered:
\begin{description}
\item[(i)]  $\Omega_{\rm m} + \Omega_\sigma + \Omega_{\Lambda_{\rm b}} \ge 0$
\ \ [Eq.~(\ref{eq:first})].
\begin{description}
\item[(a)]  $\sqrt{1 - \Omega_\kappa + \Omega_{\Lambda_{\rm b}}} -
\sqrt{\Omega_\ell} > 0$ \ $\Longrightarrow$ \ \{(\ref{hubble1}),
(\ref{omega-r1})\} and \{(\ref{hubble2}), (\ref{omega-r2})\} are possible.
\item[(b)] $\sqrt{1 - \Omega_\kappa + \Omega_{\Lambda_{\rm b}}} -
\sqrt{\Omega_\ell} < 0$ \ $\Longrightarrow$ \ only \{(\ref{hubble1}),
(\ref{omega-r1})\} is possible.
\end{description}
\item[(ii)] $\Omega_{\rm m} + \Omega_\sigma + \Omega_{\Lambda_{\rm b}} < 0$
\ \ [Eq.~(\ref{eq:second})].
\begin{description}
\item[(a)]  $\sqrt{1 - \Omega_\kappa + \Omega_{\Lambda_{\rm b}}} -
\sqrt{\Omega_\ell} > 0$ \ $\Longrightarrow$ \ only \{(\ref{hubble2}),
(\ref{omega-r2})\} is possible.
\item[(b)] $\sqrt{1 - \Omega_\kappa + \Omega_{\Lambda_{\rm b}}} -
\sqrt{\Omega_\ell} < 0$ \ $\Longrightarrow$ \ only \{(\ref{hubble1}),
(\ref{omega-r2})\} is possible.
\end{description}
\end{description}
To see these possibilities, one must consider Eqs.~(\ref{eq:intermed1}) and
(\ref{eq:intermed2}).

Let us introduce the quantity $F \equiv \sqrt{1 - \Omega_\kappa +
\Omega_{\Lambda_{\rm b}}} - \sqrt{\Omega_\ell}\,$ that enters the above
conditions (a) and (b).  Now we note that, when using (\ref{omega-r1}) for
determining $\Omega_\sigma$, we have
\begin{equation}
\Omega_{\rm m} + \Omega_\sigma + \Omega_{\Lambda_{\rm b}} = 1 - \Omega_\kappa +
\Omega_{\Lambda_{\rm b}} + 2 \sqrt{\Omega_\ell}\, \sqrt{1 - \Omega_\kappa +
\Omega_{\Lambda_{\rm b}}}\, ,
\end{equation}
so that condition (i) is satisfied automatically for any sign of $F$.
Similarly, when using (\ref{omega-r2}) for determining $\Omega_\sigma$, we have
\begin{equation}
\Omega_{\rm m} + \Omega_\sigma + \Omega_{\Lambda_{\rm b}} = 1 - \Omega_\kappa +
\Omega_{\Lambda_{\rm b}} - 2 \sqrt{\Omega_\ell}\, \sqrt{1 - \Omega_\kappa +
\Omega_{\Lambda_{\rm b}}}\, ,
\end{equation}
so that condition (ii) is satisfied automatically for $F < 0$.  This means that
conditions (i) and (ii) need not be checked in the procedure.

The case of the Randall--Sundrum constraint with $\Omega_\kappa = 0$ is much
simpler, since it allows only BRANE2 model.  Note that the condition
$\Omega_\ell \le \Omega_{\Lambda_{\rm b}}$ must be satisfied in this case. [See
the explanations above Eq.~(\ref{eq:con-rs}).] The value of $\Omega_\ell$ can
be obtained from (\ref{eq:con-rs}):
\begin{equation}
2 \sqrt{\Omega_\ell} = \frac{1 - \Omega_{\rm m}}{\sqrt{1 + \Omega_{\Lambda_{\rm
b}}} - \sqrt{\Omega_{\Lambda_{\rm b}}}} \, .
\end{equation}
So that
\begin{equation}
\Omega_\ell \le \Omega_{\Lambda_{\rm b}} \quad \Longrightarrow \quad 1 -
\Omega_{\rm m} + 2 \Omega_{\Lambda_{\rm b}} \le 2 \sqrt{\Omega_{\Lambda_{\rm
b}}} \sqrt{ 1 + \Omega_{\Lambda_{\rm b}}} \, .
\end{equation}
This inequality can be simplified by taking the square of both sides:
\begin{equation}
\left(1 - \Omega_{\rm m} + 2 \Omega_{\Lambda_{\rm b}} \right)^2 \le 4
\Omega_{\Lambda_{\rm b}} \left( 1 + \Omega_{\Lambda_{\rm b}} \right) \quad
\Longrightarrow \quad \left( 1 - \Omega_{\rm m} \right)^2 - 4 \Omega_{\rm m}
\Omega_{\Lambda_{\rm b}} \le 0 \, .
\end{equation}
Finally, we get the inequality
\begin{equation} \label{app:constraint}
\Omega_{\Lambda_{\rm b}} \ge \frac{\left(1 - \Omega_{\rm m} \right)^2}{4
\Omega_{\rm m}} \, .
\end{equation}

\bigskip

Summarizing these results, one can suggest the following procedure for testing
the braneworld models:

Specify the values of $\Omega_\ell\,$, $\Omega_{\Lambda_{\rm b}}$,
$\Omega_\kappa$, and $\Omega_{\rm m}$ (initial $\Omega$'s).  In doing this,
note that always $\Omega_\ell > 0$ by definition, $0 < \Omega_{\rm m} < 1$ from
observations, and $1 - \Omega_\kappa + \Omega_{\Lambda_{\rm b}} \ge 0$ by
physical consistency [see Eq.~(\ref{original})]. In addition, one can assume
the universe to be spatially flat: $\Omega_\kappa = 0$.

\begin{description}
\item[(A)] To study the BRANE1 model (\ref{hubble1}) with constraint
(\ref{omega-r1}):

Use (\ref{omega-r1}) to calculate $\Omega_\sigma$, and (\ref{hubble1}) to
calculate $d_L(z)$.

\item[(B)] To study the BRANE1 model (\ref{hubble1}) with constraint
(\ref{omega-r2}):

In specifying the values for the initial $\Omega$'s, always choose $\Omega_\ell
> 1 - \Omega_\kappa + \Omega_{\Lambda_{\rm b}}$.  Use (\ref{omega-r2}) to
calculate $\Omega_\sigma$, and (\ref{hubble1}) to calculate $d_L(z)$.

\item[(C)] To study the BRANE2 model (\ref{hubble2}) with constraint
(\ref{omega-r2}):

In specifying the values for the initial $\Omega$'s, always choose $\Omega_\ell
< 1 - \Omega_\kappa + \Omega_{\Lambda_{\rm b}}$. Use (\ref{omega-r2}) to
calculate $\Omega_\sigma$, and (\ref{hubble2}) to calculate $d_L(z)$.

\item[(D)] The case of {\em disappearing dark energy\/} discussed
in Sec.~\ref{sec:RS} and described by BRANE2 model is studied separately as
follows:

Set $\Omega_\kappa = 0$, specify the value of $\Omega_{\rm m}$ in the range $0<
\Omega_{\rm m} < 1$, and choose $\Omega_{\Lambda_{\rm b}}$ in the range given
by (\ref{app:constraint}).  Now use (\ref{eq:con-rs}) to find $\Omega_\ell\,$,
and then (\ref{eq:omsigma}) to find $\Omega_\sigma$.

Finally, use (\ref{hubble2}) to calculate $d_L(z)$.
\end{description}

\end{document}